\documentclass[preprint,showpacs]{revtex4}

\usepackage[latin1]{inputenc}
\usepackage{graphicx}%
\usepackage{dcolumn}
\usepackage{amsmath}

\makeatletter
\def\btt#1{\texttt{\@backslashchar#1}}%
\DeclareRobustCommand\bblash{\btt{\@backslashchar}}%
\makeatother

\begin{document}

 \title{Physics of traffic gridlock in a city  }

\mark{Physics of traffic gridlock in a city   }

\author{Boris S. Kerner $^1$ }

\affiliation{$^1$ 
Daimler AG, GR/PTF, HPC:  G021, 71059 Sindelfingen, Germany 
}



\begin{abstract}
Based of simulations of   a stochastic three-phase traffic flow model, we reveal that 
at a signalized city intersection under small link inflow rates
at which a vehicle queue developed during the red phase of light signal
dissolves fully during the   green phase, i.e., no traffic gridlock should be expected, nevertheless,
 traffic breakdown
with the subsequent city gridlock occurs  with some probability after a random time delay. This traffic breakdown is initiated by
 a first-order phase transition
  from free flow  to synchronized flow (F$\rightarrow$S transition)
occurring upstream of the vehicle queue at light signal. 
The   probability of traffic breakdown at light signal
is an increasing function of  the link inflow rate and  duration of the red phase of light signal.
 \end{abstract}

\maketitle

 \section{Introduction}

 It is known that 
at relatively great link inflow rates in a city network 
 at least some of the vehicles, which are waiting within a vehicle queue at light signal during the red phase of  light signal, 
 cannot pass the intersection during the green phase (so called an over-saturation regime of traffic). In this case,
 the queue can grow non-limited over time leading to traffic gridlock in a city.
 For this reason, through light signal control at city intersections an under-saturated 
 traffic is sought  in which 
  all vehicles, which are waiting within the queue at light signal during the red phase, 
 can pass the intersection during the green phase resulting in the fully dissolution of the queue
 during the green phase (for a review see~\cite{Gartner}). Thus no queue growth
 and no gridlock should be expected in such a city network.  
 
 Nevertheless, as we show in this article
 traffic breakdown
with the subsequent city gridlock can occur  with some probability after a random time delay.

\section{Model}
 
  To study   traffic breakdown   on city links  controlling by light signals, 
  we use the Kerner-Klenov stochastic three-phase traffic flow model for  single-lane roads that reads 
 as follows~\cite{book}:
  \begin{equation}
v_{n+1}=\max(0, \min(v_{{\rm free}}, \tilde v_{n+1}+\xi_{n}, v_{n}+a \tau, v_{{\rm s},n} )),
\label{final}
\end{equation}
\begin{equation}
\label{next_x}
x_{n+1}= x_{n}+v_{n+1}\tau,
\end{equation}
where 
 $n=0, 1, 2, ...$ is number of time steps,
$\tau=1$ sec is a time step, $x_{n}$ and $v_{n}$ are the vehicle coordinate and speed
at time step $n$,   $a$ is the maximum acceleration, $v_{\rm free}$ is a maximum speed in free flow,
$\tilde v_{n}$ is the vehicle speed  without  speed fluctuations $\xi_{n}$,
$v_{{\rm s}, n}$ is a safe speed;   model functions and parameters for vehicle motion
in a single-lane  road without light signals for a discrete model version used here
are presented in Appendix of~\cite{KKl2010}. 
  
Vehicles decelerate at the upstream front of a queue at light signal as they do this at the upstream front
of a wide moving jam propagating on a road without light signals~\cite{book}; during the green phase, vehicles accelerate at the downstream queue front with a random
time delay as they do it at the downstream jam front. During the yellow phase of light signal (between  the green and red phases)
 the vehicle  passes the light signal location, if the vehicle can do it until the end of the yellow phase; otherwise, the vehicle 
  comes to a stop at light signal.

 In all simulations presented below,  at given link inflow rates and parameters of light signals, which do not depend on time,
any initial   vehicle queue at  light signal that has been 
 developed during the red phase of light signal dissolves  fully during the green phase. 
 
 The main result of this article   is as follows:
 In this initial under-saturated traffic,  after a random time delay a first-order phase transition
  from free flow  to synchronized flow (F$\rightarrow$S transition)~\cite{book}  can occur   upstream of the queue.
  The F$\rightarrow$S transition results in traffic breakdown: the  queue at light signal 
 begins to self-grow non-reversibly leading to
 traffic gridlock in a city.
 
 \section{Features of traffic gridlock in  city}   
  
  Features of this phenomenon are as follows. 
  
\begin{figure}
\begin{center}
\includegraphics*[width=12 cm]{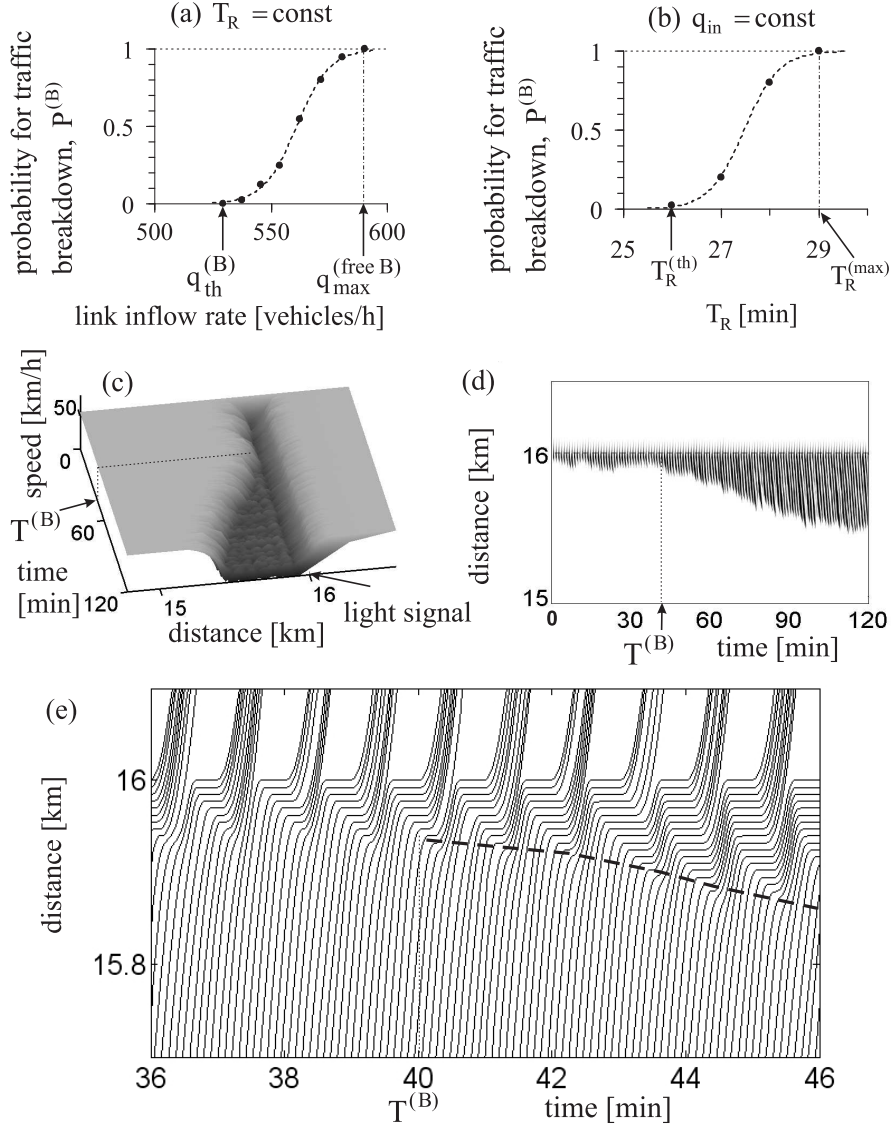}
\caption{Spontaneous traffic breakdown at light signal of city link:
 Probability of traffic breakdown $P^{\rm (B)}$  
as functions of inflow rate $q_{\rm in}$ into a  single-lane link (a) and  of duration of the red phase $T_{\rm R}$ (b). (c--e)
 Average speed in space and time (c) for simulation realization 
1, the same data as that in (c) presented by regions with variable darkness (d) (the lower the speed, the darker the region; 
in white regions the speed is higher than 15 km/h), and vehicle trajectories (e) for a time interval related to (c, d).
Cycle length of  light signal is  $\vartheta=T_{\rm G}+T_{\rm Y}+T_{\rm R}=$ 60 sec,
 duration of the yellow phase is  $T_{\rm Y}=$ 2 sec, the observation time $T_{\rm ob}=$
60 min~\cite{Realization}. In (b--e), $q_{\rm in}=$ 570 vehicles/h. In (a, c--e), duration of the red and green phases are    $T_{\rm R}=$ 28  
and  $T_{\rm G}=$ 30 sec, respectively. Light signal  is located at 16 km.    
Simulation model parameters, which are different from those presented in Table A2
 of~\cite{KKl2010}  are:    $v_{\rm free}=$ 55 km/h, $p_{a}=   0.1$,
  $a^{(\rm a)}= a$, 
  $v_{22} = 7 \ {\rm ms^{-1}}/\delta v$,  
  $\Delta v_{22} = 2 \ {\rm ms^{-1}}/\delta v$, 
$v_{01} = 6 \ {\rm ms^{-1}}/\delta v$, $v_{21} = 7 \ {\rm ms^{-1}}/\delta v$, $a=$ 0.5 ${\rm ms^{-2}}/\delta a$, where $\delta v=$ 0.01 ${\rm ms^{-1}}$, $\delta a=$ 0.01 ${\rm ms^{-2}}$. 
  \label{Main_28} } 
\end{center}
\end{figure}

  (i) There are ranges of the link inflow rate $q_{\rm in}$ (Fig.~\ref{Main_28}(a))
 and the duration of the red phase $T_{\rm R}$ of  light signal on  link (Fig.~\ref{Main_28}(b))
    \begin{eqnarray}
 \label{range1}
q^{\rm (B)}_{\rm th}\leq q_{\rm in}< q^{\rm (free \ B)}_{\rm max} \quad {\rm at} \  T_{\rm R}= {\rm const}, \\
T^{\rm (th)}_{\rm R}\leq T_{\rm R}< T^{\rm (max)}_{\rm R} \ {\rm at} \quad  q_{\rm {in}}= {\rm const}
\label{range2}
\end{eqnarray}
 within which the   under-saturation regime of 
  traffic  at light signal is in a  metastable state with respect to traffic breakdown.
  Here $q^{\rm (B)}_{\rm th}$ and $T^{\rm (th)}_{\rm R}$  are the threshold flow rate and duration of the red phase for the breakdown:
  at $q_{\rm in}<q^{\rm (B)}_{\rm th}$ or $T_{\rm R}<T^{\rm (th)}_{\rm R}$, 
  the probability of  traffic breakdown $P^{\rm (B)}$   occurring during a given observation time interval $T_{\rm ob}$~\cite{Realization} 
  is equal to zero; 
$q^{\rm (free \ B)}_{\rm max}$ and $T^{\rm (max)}_{\rm R}$ are the maximum  flow rate and duration of the red phase  
at which the breakdown probability $P^{\rm (B)}$ reaches 1.
  
  (ii) The probability of this traffic breakdown $P^{\rm (B)}$     
  is  a strongly increasing function of $q_{\rm in}$ (Fig.~\ref{Main_28}(a)) and $T_{\rm R}$  (Fig.~\ref{Main_28}(b))~\cite{Formulae}.

(iii) After the link inflow   has been switched on, traffic breakdown occurs with a time delay $T^{\rm (B)}$
 that can be considerably longer than   cycle length $\vartheta$ of light signal.
  This means that at time
  $t<T^{\rm (B)}$ the queue developed during the red phase  disolves fully during the green phase,
  i.e., the initial under-saturation regime of traffic exists at light signal.
 In particular, in Fig.~\ref{Main_28} (c--e) we can see the under-saturation regime at   light signal exists   39 cycles of  light signal (39 minutes). 
 However, at $t=T^{\rm (B)}=$ 40 min
traffic breakdown occurs spontaneously  (Fig.~\ref{Main_28} (c--e)): The average outflow rate downstream 
of the link decreases from 570 to about 535 vehicles/h~\cite{Flow} and the under-saturation regime transforms into over-saturation. 
This occurs without any change in the link inflow rate $q_{\rm in}$
and light signal parameters. This traffic breakdown results 
in the non-reversible increase in the queue length at the light signal over time (Fig.~\ref{Main_28}(c, d)). 
The beginning of this non-reversible increase in the queue length at the light signal has been labeled by a dashed curve in Fig.~\ref{Main_28}(e).

\begin{figure}
\begin{center}
\includegraphics*[width=14 cm]{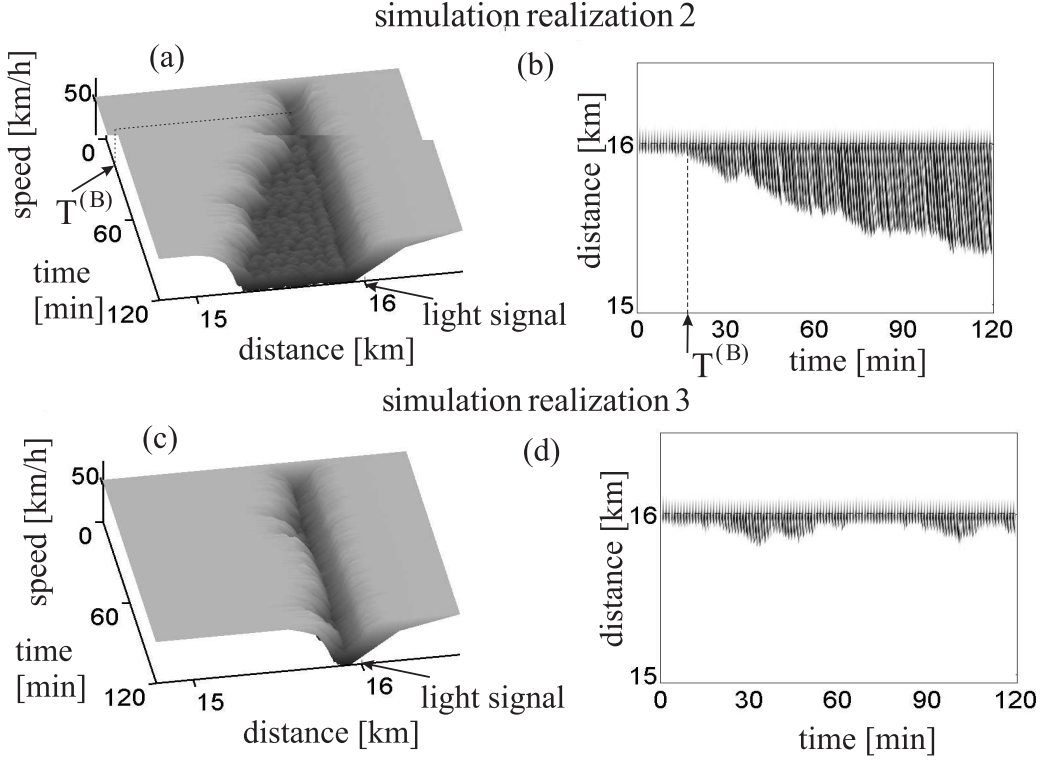}
\caption{Two different simulation realizations (runs) calculated at the same model parameters 
as those used for the simulation realization 1 shown in Fig.~\ref{Main_28} (c--e): (a, b) Traffic breakdown occurs 
after $T^{\rm (B)}=$ 20 min in realization 2. (c, d) No traffic breakdown occurs in  
realization 3. 
\label{T60_28_TB} } 
\end{center}
\end{figure}

(iv) The time delay  $T^{\rm (B)}$ of traffic breakdown at light signal
 is a random value: In different simulation realizations (runs) made 
  at the same given link inflow rate and parameters of light signal, we find very different values
  of $T^{\rm (B)}$. 
 For example, in simulation realization 1 shown in Fig.~\ref{Main_28} (c--e) as abovementioned $T^{\rm (B)}=$ 40 min, while in 
simulation realizations 2      shown in Fig.~\ref{T60_28_TB} (a, b) $T^{\rm (B)}=$   20 min. 
In some of the  realizations, no traffic breakdown occurs during    120 min (Fig.~\ref{T60_28_TB} (c, d)). 
This is associated with the result  that traffic breakdown at light signal occurs within the flow rate range 
 (\ref{range1}) with  probability  $0<P^{\rm (B)}<1$, i.e., 
 in some realizations traffic breakdown occurs, whereas in other simulation realizations (runs) 
 no traffic breakdown occurs at the same link inflow rate and light signal parameters.
 
\begin{figure}
\begin{center}
\includegraphics*[width=14 cm]{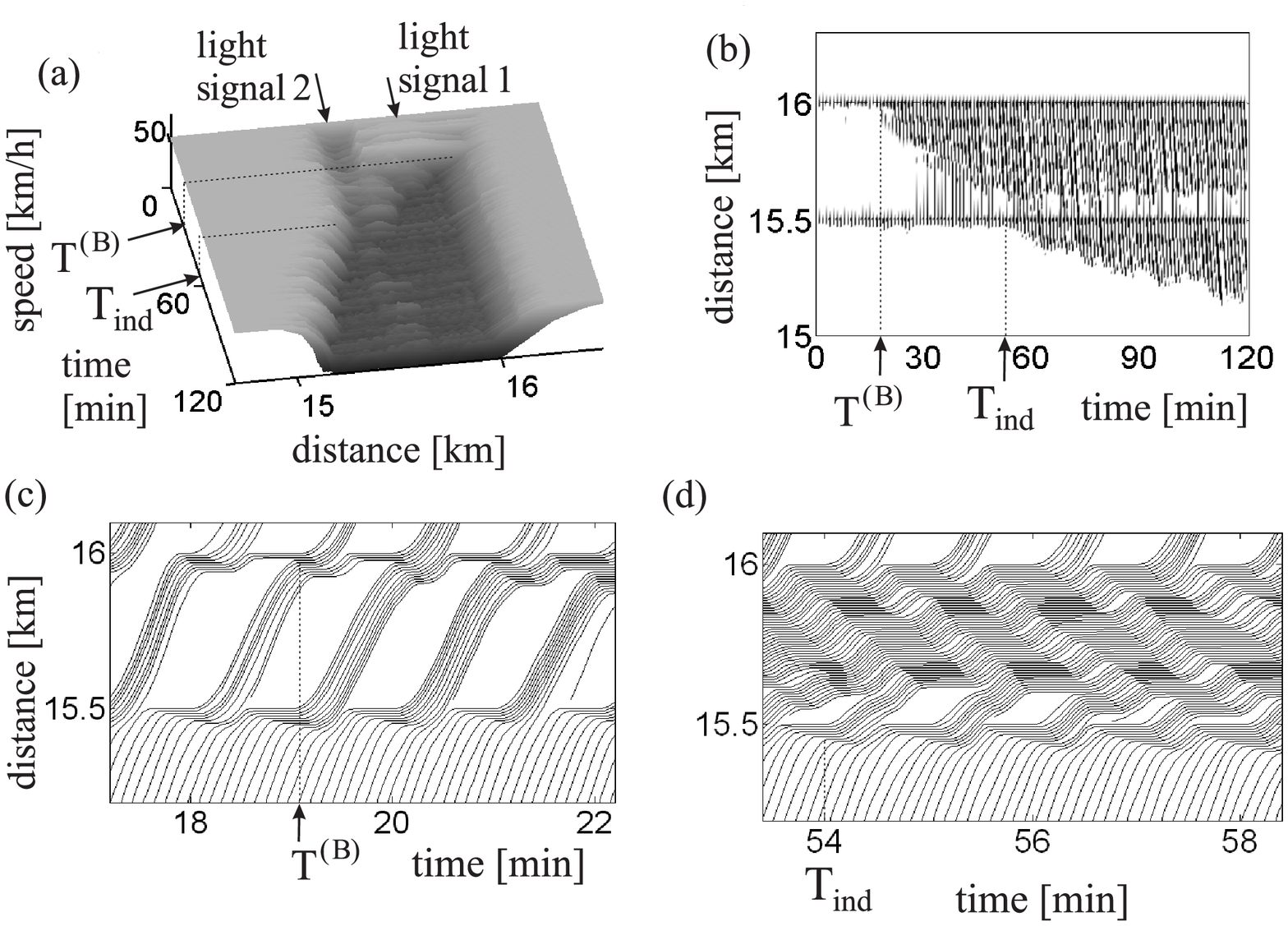}
\caption{Traffic gridlock: (a, b) Speed in space and time (a) and the same data   presented by regions with variable darkness (b). (c, d) Vehicle trajectories 
for two time intervals related to (a, b). Simulations of two  network intersections:
 Link 1 with light signal 1 (at 16 km) at the downstream intersection  is the same as that  
 in  Figs.~\ref{Main_28} and~\ref{T60_28_TB}; the inflow to link 1 consists of the outflows
 from links 2 and 3 for vehicles going to link 1 (inflow rates $q_{\rm in}=$ 545 and 36 vehicles/h for link 2 and 3, respectively) 
 during the related green phases of respective light signals 2 and 3 (located at 15.5 km) at the upstream intersection.
 After   time delay $T^{\rm (B)}=$ 19 min traffic breakdown occurs on  link 1; the growing queue induces 
 traffic breakdown  on link 2 
 at $T_{\rm ind}=$ 54 min. 
  $\vartheta=$ 60 sec,
 $T_{\rm Y}=$ 2 sec.   
  $T_{\rm R}=$ 28 for links 1 and 2,  30 sec for link 3.   
Other   model parameters are the same as those in Fig.~\ref{Main_28}.
  \label{Main2_28} } 
\end{center}
\end{figure}

 (v)  As expected, the non-reversible
 growth of the queue (Fig.~\ref{Main_28} (c--e)) causes
 gridlock in a city. Indeed, while approaching an upstream city intersection, the queue prevents the free vehicle entering
 to the link from upstream city links
 leading to traffic breakdown at the upstream intersection (Fig.~\ref{Main2_28}), and so on.

     \begin{figure}
\begin{center}
\includegraphics*[width=14 cm]{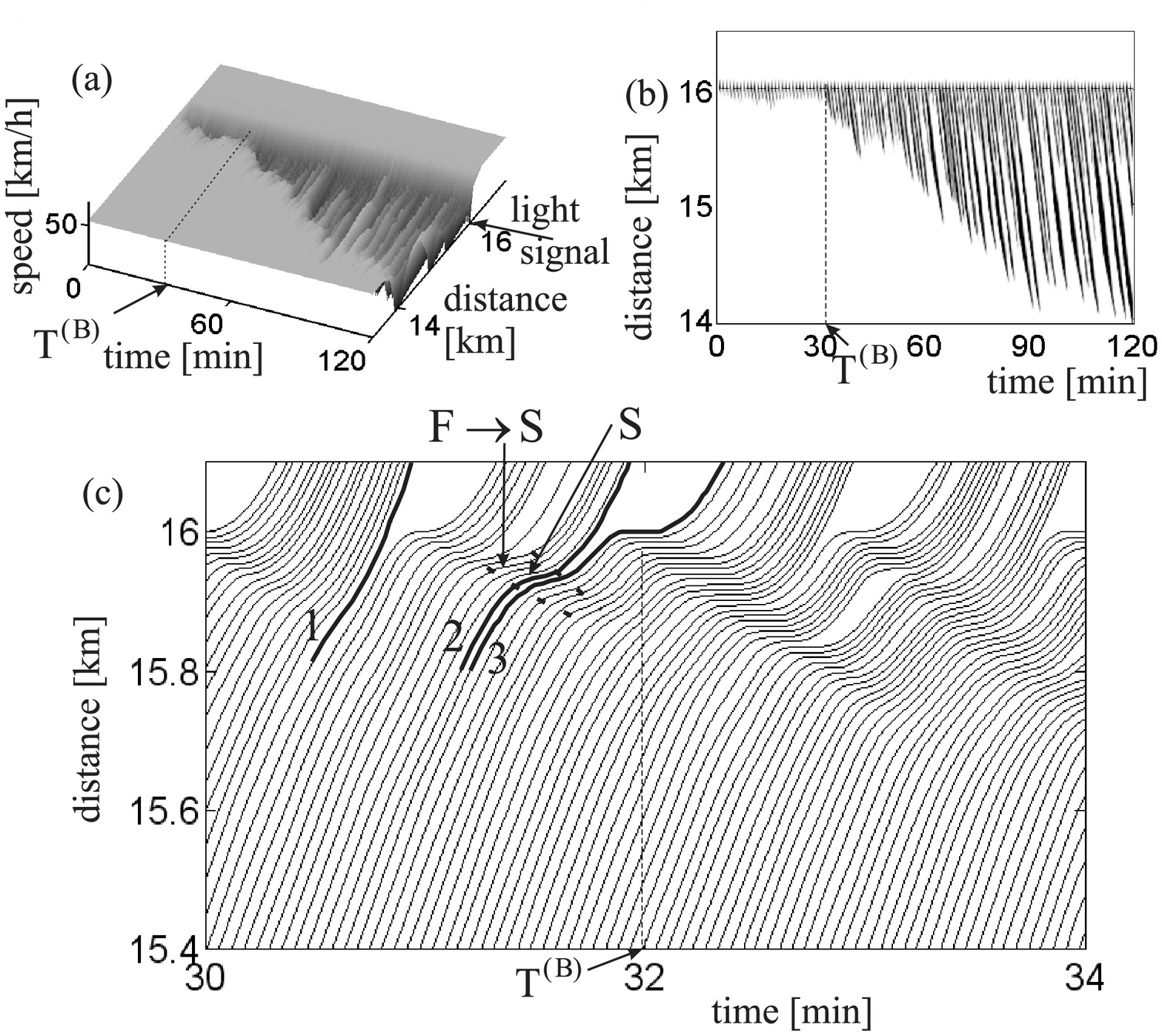}
\caption{Physics of traffic breakdown: Speed in space and time (a)  and the same data presented by regions with variable darkness (b)   
(in white regions the speed is higher than 25 km/h).  (c) Vehicle trajectories related to (a, b). 
$q_{\rm in}=$ 1060 vehicles/h.    $\vartheta=$ 60 sec,
   $T_{\rm Y}=$ 2 sec, 
   $T_{\rm R}=$ 10 sec,    $T_{\rm G}=$ 48 sec.
Other  parameters are the same as those in Fig.~\ref{Main_28}. 
\label{T60_10} } 
\end{center}
\end{figure}

  \section{The physics of  traffic breakdown at signalized city intersection}

 The physics of traffic breakdown at light signal is as follows:
 
1. At the end of the green phase in the under-saturation regime at light signal,
 there is at least one of the vehicles passing light signal without stopping 
 (see vehicle trajectories for cycles of light signal at $t\leq$ 39 min, i.e., at $t<T^{\rm (B)}$ in Fig.~\ref{Main_28}(e)). 
 Therefore, in the under-saturation regime
 the link outflow rate~\cite{Flow}  
 is greater    than after traffic breakdown has occurred leading to the over-saturation regime in which 
   the vehicle queue length   increases   non-reversibly over time (Fig.~\ref{Main_28} (c--e)). 
 
 2. Traffic breakdown at light signal is initiated by an F$\rightarrow$S transition upstream of the queue,
 as can be clearly seen if
we consider a shorter duration of the red phase $T_{\rm R}=$ 10 sec (Fig.~\ref{T60_10})
  at the same  
light signal cycle $\vartheta=$ 60 sec~\cite{Real}.  Before traffic breakdown    occurs, there are vehicles that
  pass   light signal   at the end of the green phase while 
 moving at nearly maximum free flow speed
 $v_{\rm free}$ (vehicle trajectory 1 in Fig.~\ref{T60_10} (c)). We find that just
 before traffic breakdown at $t=T^{\rm (B)}=$ 32 min occurs, an
 F$\rightarrow$S transition happens upstream of the queue developed at the previous cycle of light signal 
 $t=T^{\rm (B)}-1=$ 31 min (labeled by arrow F$\rightarrow$S in Fig.~\ref{T60_10} (c)). Due to the F$\rightarrow$S transition, 
 synchronized flow at the end of the green phase is forming (region labeled by $S$ in Fig.~\ref{T60_10} (c)).
 Thus rather than moving at a nearly free flow speed, at the end of the green phase  vehicles begin to move at a lower synchronized flow speed
 (vehicle trajectory 2  in Fig.~\ref{T60_10} (c)). As a result,   these vehicles
reach the location (16 km) of light signal later (vehicle 2) than they would do this while moving at nearly free flow speed (vehicle 1). Therefore,
a smaller number of vehicles pass light signal during the green phase
(19 and 12 vehicles for two subsequent green phases associated with vehicles 1 and 2,  respectively).
This explains why
rather than pass light signal  the following vehicle 3 in Fig.~\ref{T60_10} (c)
must stop during the next red phase.
 Thus the F$\rightarrow$S transition
at the previous cycle of light signal results in traffic breakdown, i.e., in the emergence of the growing   queue because
due to a lower speed in synchronized flow the smaller number of vehicles pass light signal during the previous green phase.

  \begin{figure}
\begin{center}
\includegraphics*[width=8 cm]{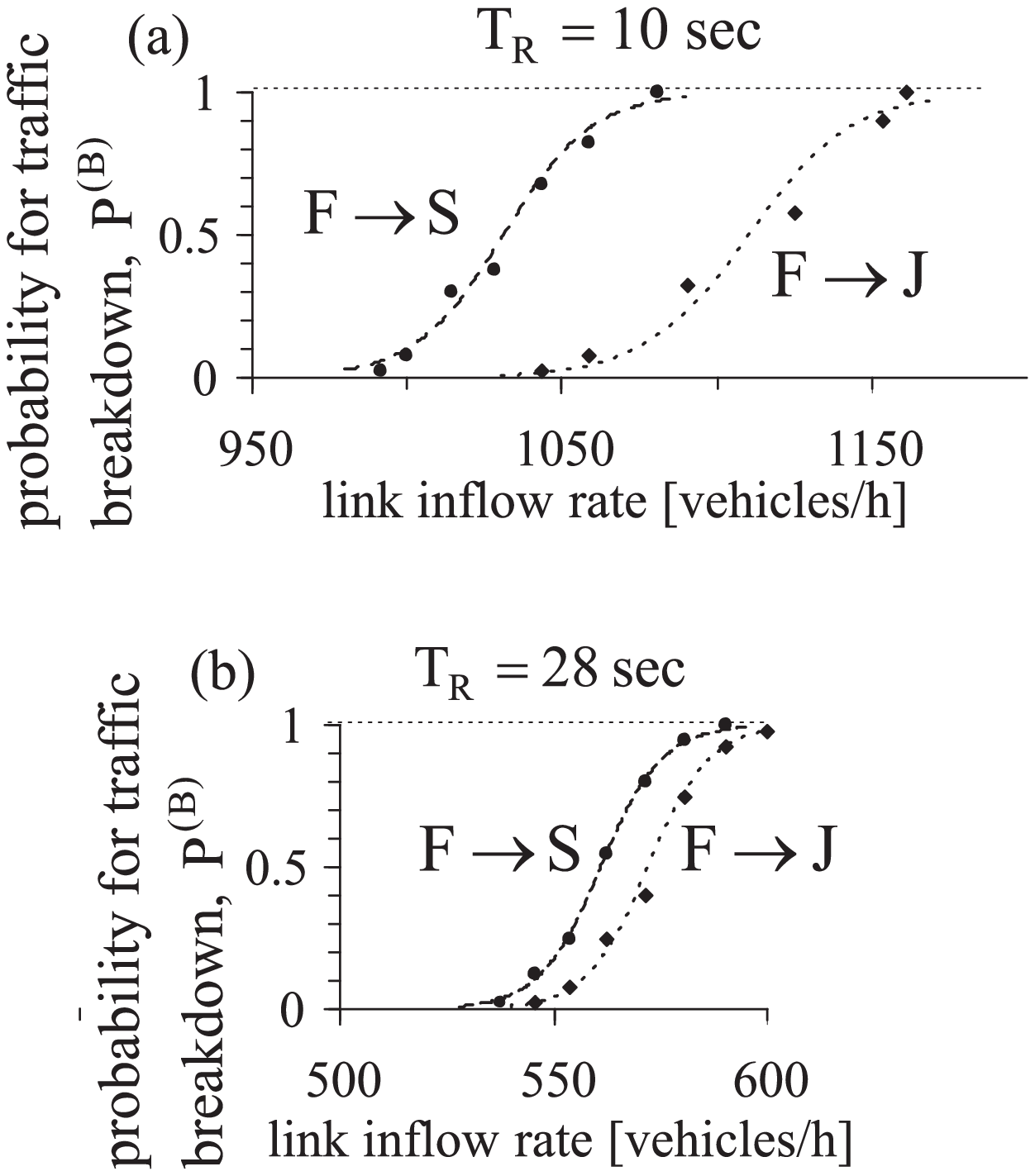}
\caption{Comparison of  the probability of traffic breakdown initiated by F$\rightarrow$S transition (curves labeled by F$\rightarrow$S)
with the probability of traffic breakdown through a direct F$\rightarrow$J transition (curves labeled by F$\rightarrow$J) for $(T_{\rm R}, \ T_{\rm G}, \ T_{\rm Y})=$ (10, 48, 2) (a)
and (28, 30, 2) (b) sec.  $\vartheta=$ 60 sec. Curve labeled by F$\rightarrow$S in (b) is taken from Fig.~\ref{Main_28} (a).
A two-phase  model used for simulations of  F$\rightarrow$J transition  follows from the Kerner-Klenov three-phase model after 
removing   the description of driver behaviors  associated with three-phase theory~\cite{book} --
  2D-region of synchronized flow states and the associated speed adaptation effect within these states as well as the over-acceleration effect
  have been removed; this 
  is done through the use of   $G_{n}= 0$ and   $p_{a}=p_{1}=p_{2}=0$ in Table A1 of~\cite{KKl2010}
  for the three-phase model. 
  All   characteristics of the F$\rightarrow$J transition and resulting wide moving jams
  in the three-phase and 
  two-phase models are identical, in particular, the flow rate in free flow (at maximum speed 55 km/h) in the outflow from a wide moving jam
  is equal to $q_{\rm out}=$ 1614 vehicles/h.
Other model parameters are the same as those in Fig.~\ref{Main_28}. 
\label{FS_FJ} } 
\end{center}
\end{figure}

3. There is  a great difference between highway traffic  
 and traffic at light signal:
 In contrast with a highway bottleneck,
light signal
 introduces always  a very great non-homogeneity in traffic while forcing vehicles to stop during the red phase. 
 However, there is a common feature of traffic breakdown at a highway bottleneck and at
light signal.  The probability of the emergence of a wide moving jam (J)
 at the highway bottleneck through    a sequence of F$\rightarrow$S$\rightarrow$J transitions is greater than
 the probability of the jam formation in free flow (F$\rightarrow$J transition)~\cite{book}. Qualitatively the same result has been found in this article: 
 Traffic breakdown at light signal, i.e., the growing queue formation
 initiated by an F$\rightarrow$S transition (Figs.~\ref{Main_28}--\ref{T60_10}) can be considered a sequence of F$\rightarrow$S$\rightarrow$J transitions
 (in contrast, the growing queue formation at light signal
without an F$\rightarrow$S transition can   be considered 
an F$\rightarrow$J transition).
 The probability of this sequence of F$\rightarrow$S$\rightarrow$J transitions at light signal
is greater than the probability for the breakdown    through an F$\rightarrow$J transition.  
  To show this, we repeat the above simulations  with
a stochastic {\it two-phase} traffic model 
  (see caption of Fig.~\ref{FS_FJ}). As in other two-phase models (see reviews~\cite{Reviews}), 
  in the two-phase model there is {\it no} a first-order   F$\rightarrow$S transition 
and, therefore,  traffic breakdown  is caused by an F$\rightarrow$J transition.
  Simulations of the three-phase and two-phase models
  show that for main city links   for which  the green phase duration is not appreciably shorter than the red phase duration~\cite{Equal}, 
 at any link inflow rate the probability for traffic breakdown initiated by the F$\rightarrow$S transition is considerably greater 
 than for the F$\rightarrow$J transition (Fig.~\ref{FS_FJ}). 
 
  \section{Conclusions}

We have found that in an initial under-saturated traffic controlled by light signal
  in which a vehicle queue developed during the red phase of light signal
dissolves fully during the   green phase, i.e., no traffic gridlock should be expected, nevertheless, traffic breakdown leading to city gridlock  
 can occur with some probability after a random time delay.
 This traffic breakdown is initiated  by an F$\rightarrow$S transition occurring upstream of the vehicle queue at light signal.
The   probability of traffic breakdown at light signal
is an increasing function of  the duration of the red phase of light signal and link inflow rate. A similar flow rate dependence of the breakdown probability  
has been found for a highway bottleneck~\cite{book}; in this sense,
light signal can be considered a bottleneck in a city network.
Therefore,   for the dynamic  control of a city network that
  could be a very interesting task for future investigations
the breakdown minimization (BM) principle   can be applied~\cite{BM,Principle}.

  I thank S. Klenov and V. Friesen for discussions  and S. Klenov for help in   simulations.


\begin{thebibliography}{8.}
\bibitem{Realization} 
To find the probability of traffic breakdown
 $P^{\rm (B)}$, a study of traffic breakdown at light signal  during a observation time $T_{\rm ob}$
 is repeated for 40 different realizations for  the same given  link flow rates and light signal control parameters, however,
 at different initial conditions
 for random model fluctuations.
\bibitem{Formulae} 
 
 These functions
   can be approximated by formulae:
$P^{(\rm B)}(q_{\rm in})=[1+\exp[\beta (q_{\rm p}-q_{\rm in})] ]^{-1}$,
$P^{(\rm B)}(T_{\rm R})=[1+\exp[\beta_{\rm R} (T_{\rm p}-T_{\rm R})] ]^{-1}$,
where $\beta$ and $q_{\rm p}$ depends on parameters of light signal control, whereas $\beta_{\rm R}$ and $T_{\rm p}$   depend on  the link inflow rate $q_{\rm in}$
and  cycle length    $\vartheta$.
\bibitem{Flow}
 The link outflow rate is averaged during time that is considerably longer than   cycle length $\vartheta$ of light signal.
\bibitem{Real} 
  
  This relation between green and red phases of light signal corresponds to a realistic case in which the link  is related to a main road
  at a city intersection.
\bibitem{Equal}
  For $\vartheta=$ 60 sec and $T_{\rm Y}=$ 2 sec (Figs.~\ref{Main_28}--\ref{FS_FJ}), the probabilities
  for traffic breakdown through the F$\rightarrow$S$\rightarrow$J and F$\rightarrow$J transitions
  become to be identical only at $T_{\rm R}\geq$ 37 sec.
\end{thebibliography}
\end{document}